\documentclass[12pt,journal,draft,onecolumn]{IEEEtran}
\usepackage{ifpdf}

\usepackage{cite}

\ifCLASSINFOpdf
\else
\fi
\usepackage[cmex10]{amsmath}
\usepackage{amssymb}
\usepackage{graphicx}
\usepackage{array}

\newtheorem{lemma}{Lemma}

\begin{document}
%
\title{Gaussian Assumption: the Least Favorable but the Most Useful}
%
%
%

\author{Sangwoo~Park, Erchin~Serpedin, and Khalid~Qaraqe}

\maketitle



%
\IEEEpeerreviewmaketitle

\IEEEPARstart{G}{aussian} assumption is the most well-known and widely used distribution in many  fields such as engineering, statistics and physics. One of the major reasons why the Gaussian distribution has become so prominent is because of the Central Limit Theorem (CLT) and the fact that the distribution of  noise in numerous engineering systems is well captured by the Gaussian distribution. Moreover, features such as analytical tractability and easy generation of other distributions from the Gaussian distribution contributed further to the popularity of Gaussian distribution. Especially, when there is no information about the distribution of observations,  Gaussian assumption appears as the most conservative choice. This follows from the fact that the Gaussian distribution minimizes the Fisher information, which is the inverse of the Cram\'{e}r-Rao lower bound (CRLB) (or equivalently stated, the Gaussian distribution maximizes the CRLB). Therefore, any optimization based on the CRLB under the Gaussian assumption can be considered to be min-max optimal in the sense of minimizing the largest CRLB (see \cite{LargeCRB:Stoica} and the references cited therein).

Inspired by the early isoperimetric inequality for entropy introduced by Costa and Cover \cite{EPI_BrunnInequality:Costa} and the more recent results of Rioul \cite{EPI:Rioul},  Stoica and Babu \cite{LargeCRB:Stoica}, the goals of this paper are threefold: i) to illustrate a connection between  \cite{LargeCRB:Stoica} and the recent information theoretic results reported in \cite{EPI_BrunnInequality:Costa}, \cite{EPI:Rioul}, ii) to present information theoretic and estimation theoretic justifications for the fact that the Gaussian assumption leads to the largest CRLB, iii) to show a slight extension of this result to the more general framework of correlated observations. Even though Stoica and Babu provided a simple and quite general proof of result that the largest CRLB is achievable by the Gaussian distribution, the proposed proof  is only applicable to the situation when  the observations are independent, i.e., the observation noise is white \cite{LargeCRB:Stoica}. However, this result can be generalized to arbitrary correlations among samples. In many practical circumstances, the correlation of the noise is inevitable since the observed data comes from a filter, and the filter introduces correlation. Therefore, the importance of this generalization cannot be ignored. This result is also closely related to two well-known results in information theory: first, the fact that a Gaussian random vector maximizes a differential entropy, and second, the worst additive noise lemma (see \cite{EPI:Rioul}, \cite{WAN:Diggavi}, and the references cited therein). Several researchers have  investigated  relationships
between estimation theoretic (statistical) concepts such as mean-square error and Fisher information and information theoretic concepts such as entropy and mutual information  (see e.g., \cite{EPI_BrunnInequality:Costa}, \cite{EPI:Rioul} and the references cited therein). However, most of these results are inclined to be rather theoretical   than practical. In this paper, we show how some of these results can be adopted to a more practical application involving the estimation
of a communication channel via a training sequence.

\section{RELEVANCE}

The approach  introduced herein paper can be adapted to optimally estimate unknown (deterministic or random) parameters  in  additive noise channels. As presented in the channel model (\ref{eq1_1}), the additive noise channel is  very general in the sense that the only assumption is the independence between data $\mathbf{x}_{\boldsymbol{\theta}}$ and noise $\mathbf{w}$. Namely, the channel model does not require the  Gaussian noise assumption, it admits  correlation among noise terms, and it also allows for correlation among data terms. Therefore, the proposed approach can be generally used in  signal processing applications involving parameter estimation, spectrum estimation, and optimization, wireless communications and information theory. This lecture note is also beneficial to courses related to  such topics.

\section{PREREQUISITES}

The readers may require some knowledge about linear algebra, elementary probability theory, statistical signal processing, and basic information theory.

\section{PROBLEM STATEMENT}

Consider a random vector $\mathbf{y}$ which is generated by the following system of equations:
\begin{eqnarray}
\label{eq1_1}   \mathbf{y} & = & \mathbf{x}_{\scriptscriptstyle\boldsymbol{\theta}} + \mathbf{w} ,
\end{eqnarray}
where $\mathbf{y}$ is an  $n \times 1$ observed random vector, $\mathbf{x}_{\scriptscriptstyle\boldsymbol{\theta}}$ denotes an $n\times 1$ signal (random) vector which depends on a $k \times 1$ unknown deterministic parameter vector $\boldsymbol{\theta}$, and $\mathbf{w}$ stands for the $n\times 1$ zero-mean noise  vector whose covariance matrix is $\boldsymbol{\Sigma}_{\mathbf{w}}$.   Random vectors $\mathbf{x}_{\boldsymbol{\theta}}$ and $\mathbf{w}$ are assumed independent of each other. The systems represented by the channel model (\ref{eq1_1}) are quite numerous. In particular,  the channel model (\ref{eq1_1}) might consist of the samples of an arbitrary stochastic process such as ARMA (autoregressive moving average) or  ARMAX (ARMA with eXogenous inputs), as mentioned in \cite{LargeCRB:Stoica}.

Based on the channel model (\ref{eq1_1}), we define the score function:
\begin{eqnarray}
\label{eq2_1}   \mathbf{s}(\boldsymbol{\theta}) & = & \nabla_{\scriptscriptstyle\boldsymbol{\theta}} \log f_{\mathbf{y}|\mathbf{x}_{\scriptscriptstyle\boldsymbol{\theta}}}(\mathbf{y}|\mathbf{x}_{\scriptscriptstyle\boldsymbol{\theta}}),
\end{eqnarray}
where $\nabla_{\scriptscriptstyle\boldsymbol{\theta}}$ denotes the  gradient with respect to $\boldsymbol{\theta}$, and $f_{\mathbf{y}|\mathbf{x}_{\scriptscriptstyle\boldsymbol{\theta}}}(\mathbf{y}|\mathbf{x}_{\scriptscriptstyle\boldsymbol{\theta}})$ is the  conditional density function of $\mathbf{y}$ given $\mathbf{x}_{\scriptscriptstyle{\boldsymbol{\theta}}}$.
The Cram\'{e}r-Rao lower bound (CRLB) is expressed by the diagonal elements of the inverse of the Fisher information matrix (FIM), and the FIM is represented as:
\begin{eqnarray}
\label{eq3_1}   \mathbf{J}_{\scriptscriptstyle\boldsymbol{\theta}} (\mathbf{y}) & = & \mathbb{E}_{\mathbf{y}} [ \mathbf{s} (\boldsymbol{\theta})  \mathbf{s}(\boldsymbol{\theta})^{T} ],
\end{eqnarray}
where the notation $\mathbb{E}_{\mathbf{y}}[\cdot]$  stands for  the  expectation with respect to a random vector $\mathbf{y}$,  and superscript $T$ denotes the operation of transposition for a vector or matrix.

Our goal is to find an optimal estimator for the parameter $\boldsymbol{\theta}$ in the sense that the estimated parameter minimizes the lower bound of the mean square error of the estimator in the worst case scenario.

\section{MINIMUM FISHER INFORMATION-A STATISTICAL VIEWPOINT}

One of the common  approaches to estimate unknown parameters is to build estimators that minimize the Cramer-Rao lower bound.  
 Since  CRLB  is expressed as the inverse of  FIM, minimizing the Cram\'{e}r-Rao lower bound is equivalent to maximizing  FIM. Given the channel model (\ref{eq1_1}), the score function in (\ref{eq2_1}) and the FIM in (\ref{eq3_1}) can be re-expressed by the following procedure.

Since $f_{\mathbf{y}|\mathbf{x}_{\scriptscriptstyle\boldsymbol{\theta}}}(\mathbf{y}|\mathbf{x}_{\scriptscriptstyle\boldsymbol{\theta}}) = f_{\mathbf{w}}(\mathbf{w})\big|_{\mathbf{w}=\mathbf{y}-\mathbf{x}_{\scriptscriptstyle\boldsymbol{\theta}}}= f_{\mathbf{w}}(\mathbf{y}-\mathbf{x}_{\scriptscriptstyle\boldsymbol{\theta}})$,  where $f_{\mathbf{w}}(\cdot)$ denotes the density function of the noise $\mathbf{w}$, and $\mathbf{x}_{\scriptscriptstyle\boldsymbol{\theta}}$ and $\mathbf{w}$ are independent of each other, using the chain rule for computing the derivative of a function, the score function $\mathbf{s}(\boldsymbol{\theta})$ is re-written as:
\begin{eqnarray}
\label{eq4_1}   \mathbf{s}(\boldsymbol{\theta}) & = & \nabla_{\scriptscriptstyle\boldsymbol{\theta}} \log f_{\mathbf{y}|\mathbf{x}_{\scriptscriptstyle\boldsymbol{\theta}}}(\mathbf{y}|\mathbf{x}_{\scriptscriptstyle\boldsymbol{\theta}})\nonumber\\
& = & \nabla_{\scriptscriptstyle\boldsymbol{\theta}} \log f_{\mathbf{w}}(\mathbf{y}-\mathbf{x}_{\scriptscriptstyle\boldsymbol{\theta}})\nonumber\\
& = & -\nabla_{\scriptscriptstyle\boldsymbol{\theta}} \mathbf{x}_{\scriptscriptstyle\boldsymbol{\theta}} \nabla_{\mathbf{w}} \log f_{\mathbf{w}}(\mathbf{w}),
\end{eqnarray}
where the gradient (Jacobian)  of the vector $\mathbf{x}_{\scriptscriptstyle\boldsymbol{\theta}}$ is defined as the $k\times n$ matrix
$\nabla_{\scriptscriptstyle\boldsymbol{\theta}} \mathbf{x}_{\scriptscriptstyle\boldsymbol{\theta}}$ with its $(i,j)th$ entry equal to
$\frac{\partial x_{\theta,j}}{\partial \theta_i}$.
Now it turns out that the FIM (\ref{eq3_1}) can be  expressed as:
\begin{eqnarray}
\label{eq5_1}   \mathbf{J}_{\scriptscriptstyle\boldsymbol{\theta}} (\mathbf{y}) & = & \mathbb{E}_{\mathbf{x}_{\scriptscriptstyle\boldsymbol{\theta}},\mathbf{w}} \left[\left(\nabla_{\scriptscriptstyle\boldsymbol{\theta}} \mathbf{x}_{\scriptscriptstyle\boldsymbol{\theta}}  \nabla_{\mathbf{w}} \log f_{\mathbf{w}}(\mathbf{w})\right) \left(\nabla_{\scriptscriptstyle\boldsymbol{\theta}} \mathbf{x}_{\scriptscriptstyle\boldsymbol{\theta}} \nabla_{\mathbf{w}} \log f_{\mathbf{w}}(\mathbf{w})\right)^T\right]\nonumber\\
\label{eq5_2}   &= & \mathbb{E}_{\mathbf{x}_{\scriptscriptstyle\boldsymbol{\theta}},\mathbf{w}} \left[\nabla_{\scriptscriptstyle\boldsymbol{\theta}} \mathbf{x}_{\scriptscriptstyle\boldsymbol{\theta}} \left(\nabla_{\mathbf{w}} \log f_{\mathbf{w}}(\mathbf{w})\nabla_{\mathbf{w}} \log f_{\mathbf{w}}(\mathbf{w})^T  \right) \nabla_{\scriptscriptstyle\boldsymbol{\theta}} \mathbf{x}_{\scriptscriptstyle\boldsymbol{\theta}}^T \right]\\
\label{eq5_3}   &= & \mathbb{E}_{\mathbf{x}_{\scriptscriptstyle\boldsymbol{\theta}}} \left[\nabla_{\scriptscriptstyle\boldsymbol{\theta}} \mathbf{x}_{\scriptscriptstyle\boldsymbol{\theta}} \mathbf{J}(\mathbf{w})  \nabla_{\scriptscriptstyle\boldsymbol{\theta}} \mathbf{x}_{\scriptscriptstyle\boldsymbol{\theta}}^T \right],
\end{eqnarray}
where the FIM with respect to $\mathbf{w}$ is defined as
\begin{eqnarray}
\label{eq5_4}   \mathbf{J}(\mathbf{w}) & = & \mathbb{E}_{\mathbf{w}} \left[\nabla_{\mathbf{w}} \log f_{\mathbf{w}}(\mathbf{w})\nabla_{\mathbf{w}} \log f_{\mathbf{w}}(\mathbf{w})^T\right].
\end{eqnarray}
In equation (\ref{eq5_2}), the expectation with respect to both $\mathbf{x}_{\scriptscriptstyle\boldsymbol{\theta}}$ and $\mathbf{w}$ can be separated into the outer expectation with respect to $\mathbf{x}_{\scriptscriptstyle\boldsymbol{\theta}}$ and the inner expectation with respect to $\mathbf{w}$ since $\mathbf{x}_{\scriptscriptstyle\boldsymbol{\theta}}$ and $\mathbf{w}$ are independent of each other. When the vector $\mathbf{x}_{\scriptscriptstyle\boldsymbol{\theta}}$ is deterministic, the outer expectation is not required. Therefore, the term related to the random vector $\mathbf{w}$ becomes the FIM, $\mathbf{J}(\mathbf{w})$, defined in equation (\ref{eq5_4}), and it is not affected by the outer expectation $\mathbb{E}_{\mathbf{x}_{\scriptscriptstyle\boldsymbol{\theta}}}[\cdot]$ in equation (\ref{eq5_3}).

The following result states that the FIM $\mathbf{J}(\mathbf{w})$, which is a positive semi-definite matrix, is lower-bounded by the  FIM $\mathbf{J}(\mathbf{w}_{\scriptscriptstyle G})$ of a normally distributed random vector $(\mathbf{w}_{\scriptscriptstyle G})$.
\begin{lemma}[Cram\'{e}r-Rao Inequality]\label{lem1}
For a random vector $\mathbf{w}$ and a Gaussian random vector $\mathbf{w}_{\scriptscriptstyle G}$ whose covariance matrix $\boldsymbol{\Sigma}_{\mathbf{w}}$ is identical to the covariance matrix of $\mathbf{w}$, the following inequality is satisfied:
\begin{eqnarray}
\label{eq6_1}   \mathbf{J}(\mathbf{w}) &\succeq& \mathbf{J}(\mathbf{w}_{\scriptscriptstyle G}),\nonumber
\end{eqnarray}
where notation $\succeq$ stands for  ``greater than or equal to'',  in the sense of the partial ordering of positive semi-definite matrices.
\begin{proof}
The  proof follows essentially  \cite{EPI:Rioul}. First, we define the following two score functions:
\begin{eqnarray}
\label{eq7_1}   \mathbf{s}_{\mathbf{w}}(\mathbf{w}) & = & \nabla_{\mathbf{w}} \log f_{\mathbf{w}}(\mathbf{w}),\nonumber\\
\mathbf{s}_{\mathbf{w}_{\scriptscriptstyle G}}(\mathbf{w}) & = & \nabla_{\mathbf{w}} \log f_{\mathbf{w}_{\scriptscriptstyle G}}(\mathbf{w}).
\end{eqnarray}
The covariance matrix of the difference of the two score functions (\ref{eq7_1}) is expressed as
\begin{eqnarray}
\label{eq8_1}   \mathbb{E}_{\mathbf{w}}\left[\left(\mathbf{s}_{\mathbf{w}}(\mathbf{w})-\mathbf{s}_{\mathbf{w}_{\scriptscriptstyle G}}(\mathbf{w})\right)\left(\mathbf{s}_{\mathbf{w}}(\mathbf{w})-\mathbf{s}_{\mathbf{w}_{\scriptscriptstyle G}}(\mathbf{w})\right)^T\right],
\end{eqnarray}
and it is always greater than or equal to the zero matrix $\mathbf{0}$ in terms of the positive semi-definite partial ordering.
Notice further that  (\ref{eq8_1}) can be simplified to
\begin{eqnarray}
\label{eq9_1}   &&\mathbb{E}_{\mathbf{w}}\left[\left(\mathbf{s}_{\mathbf{w}}(\mathbf{w})-\mathbf{s}_{\mathbf{w}_{\scriptscriptstyle G}}(\mathbf{w})\right)\left(\mathbf{s}_{\mathbf{w}}(\mathbf{w})-\mathbf{s}_{\mathbf{w}_{\scriptscriptstyle G}}(\mathbf{w})\right)^T\right] \nonumber\\
& = & \mathbf{J}(\mathbf{w}) -\mathbb{E}_{\mathbf{w}}\left[\mathbf{s}_{\mathbf{w}}(\mathbf{w})\mathbf{s}_{\mathbf{w}_{\scriptscriptstyle G}}(\mathbf{w})^T\right]-\mathbb{E}_{\mathbf{w}}\left[\mathbf{s}_{\mathbf{w}_{\scriptscriptstyle G}}(\mathbf{w})\mathbf{s}_{\mathbf{w}}(\mathbf{w})\right]+ \mathbf{J}(\mathbf{w}_{\scriptscriptstyle G})\nonumber\\
& = &  \mathbf{J}(\mathbf{w}) - \mathbf{J}(\mathbf{w}_{\scriptscriptstyle G}).
\end{eqnarray}
Since $\mathbf{w}_{\scriptscriptstyle G}$ is a Gaussian random vector, $\mathbf{s}_{\mathbf{w}_{\scriptscriptstyle G}}(\mathbf{w}) = -\boldsymbol{\Sigma}_{\mathbf{w}}^{-1}\mathbf{w}$. Also, $\mathbb{E}_{\mathbf{w}}\left[\mathbf{s}_{\mathbf{w}}(\mathbf{w})\mathbf{s}_{\mathbf{w}_{\scriptscriptstyle G}}(\mathbf{w})^T \right] = - \int  \left(\nabla_{\mathbf{w}} f_{\mathbf{w}}(\mathbf{w})\right)\mathbf{w}^T d\mathbf{w} \boldsymbol{\Sigma}_{\mathbf{w}}^{-1}= \int  f_{\mathbf{w}}(\mathbf{w}) d\mathbf{w} \boldsymbol{\Sigma}_{\mathbf{w}}^{-1}=\boldsymbol{\Sigma}_{\mathbf{w}}^{-1}$ by Green's identity (see e.g., \cite{EPI_BrunnInequality:Costa} and the references cited therein). Here, Green's identity plays the role of the integration by parts for a vector. Since $\mathbf{J}(\mathbf{w}_{\scriptscriptstyle G})=\boldsymbol{\Sigma}_{\mathbf{w}}^{-1}$, the last equality in equation (\ref{eq9_1}) is verified.
Since the covariance matrix is always  positive semi-definite, from equation (\ref{eq9_1}),
\begin{eqnarray}
\label{eq10_1}  \mathbb{E}_{\mathbf{w}}\left[\left(\mathbf{s}_{\mathbf{w}}(\mathbf{w})-\mathbf{s}_{\mathbf{w}_{\scriptscriptstyle G}}(\mathbf{w})\right)\left(\mathbf{s}_{\mathbf{w}}(\mathbf{w})-\mathbf{s}_{\mathbf{w}_{\scriptscriptstyle G}}(\mathbf{w})\right)^T\right] & = &  \mathbf{J}(\mathbf{w}) - \mathbf{J}(\mathbf{w}_{\scriptscriptstyle G}) \succeq \mathbf{0}.
\end{eqnarray}
Therefore, the proof is completed.
\end{proof}
\end{lemma}

Due to  Lemma \ref{lem1}, when $\mathbf{w}$ is a Gaussian random vector, the FIM $\mathbf{J}(\mathbf{w})$ is minimized, and consequently the FIM $\mathbf{J}_{\boldsymbol{\theta}}(\mathbf{y})$ is also minimized:
\begin{eqnarray}
\label{eq12_1}  \mathbf{J}_{\theta} (\mathbf{y}) & = & \mathbb{E}_{\mathbf{x}_{\scriptscriptstyle\boldsymbol{\theta}}} \left[\nabla_{\scriptscriptstyle\boldsymbol{\theta}} \mathbf{x}_{\scriptscriptstyle\boldsymbol{\theta}} \mathbf{J}(\mathbf{w})  \nabla_{\scriptscriptstyle\boldsymbol{\theta}} \mathbf{x}_{\scriptscriptstyle\boldsymbol{\theta}}^T \right]\nonumber\\
& \succeq & \mathbb{E}_{\mathbf{x}_{\scriptscriptstyle\boldsymbol{\theta}}} \left[\nabla_{\scriptscriptstyle\boldsymbol{\theta}} \mathbf{x}_{\scriptscriptstyle\boldsymbol{\theta}}  \mathbf{J}(\mathbf{w}_{\scriptscriptstyle G}) \nabla_{\scriptscriptstyle\boldsymbol{\theta}} \mathbf{x}_{\scriptscriptstyle\boldsymbol{\theta}}^T \right]\nonumber\\
& = & \mathbf{J}_{\theta}(\bar{\mathbf{y}}),
\end{eqnarray}
where $\bar{\mathbf{y}}=\mathbf{x}_{\scriptscriptstyle \boldsymbol{\theta}}+\mathbf{w}_{\scriptscriptstyle G}$, and the equality holds if and only if $\mathbf{w}$ is normally distributed.
The inequality in equation (\ref{eq12_1}) is due to the fact that  for an arbitrary matrix $\mathbf{C}$, the inequality $\mathbf{C}\mathbf{A}\mathbf{C}^T \preceq \mathbf{C}\mathbf{B}\mathbf{C}^T$ holds whenever positive semi-definite matrices $\mathbf{A}$ and $\mathbf{B}$ satisfy $\mathbf{A} \preceq \mathbf{B}$.

From equations (\ref{eq5_3}) and (\ref{eq12_1}), we know that the CRLB depends on the parameter $\mathbf{\theta}$ only through the FIM, $\mathbf{J}(\mathbf{w})$. In other words, the CRLB only depends on $\mathbf{J}(\mathbf{w})$ when $\mathbf{x}_{\theta}$ is fixed. Therefore, the Gaussian random vector $\mathbf{w}_{\scriptscriptstyle G}$ maximizes the CRLB (or, equivalently minimizes the FIM, $\mathbf{J}_{\boldsymbol{\theta}}(\mathbf{y})$), when $\mathbf{x}_{\scriptscriptstyle\boldsymbol{\theta}}$ is fixed.
Therefore, any design which optimizes the FIM (\ref{eq5_3}) (or equivalently the CRLB) when the random vector $\mathbf{w}$ is Gaussian, can be considered min-max optimal in the light of generating the smallest FIM (or the largest CRLB) in the worst  situation.

\section{MINIMUM MUTUAL INFORMATION-AN INFORMATION THEORETIC VIEWPOINT}

It is well-known that, given the covariance matrix, a Gaussian random vector minimizes the FIM, a result referred to as the  Cram\'{e}r-Rao inequality (see \cite{LargeCRB:Stoica}, \cite{EPI:Rioul}, and the references cited therein). On the other hand, a Gaussian random vector maximizes a differential entropy when the covariance matrix is given (see \cite{EPI:Rioul}, \cite{inf:cover}, and the references cited therein). These two results are closely related to each other.  First, consider  this relationship for random variables. Given a random variable $w$ and a Gaussian random variable $w_{\scriptscriptstyle G}$, the following inequalities are satisfied:
\begin{itemize}
\item $J(w) \geq J(w_{\scriptscriptstyle G})$ when $N(w)=N(w_{\scriptscriptstyle G})$,
\item $N(w) \geq N(w_{\scriptscriptstyle G})$ when $J(w)=J(w_{\scriptscriptstyle G})$,
\end{itemize}
where $N(\cdot)$ denotes the entropy power of a random variable, and $J(\cdot)$  stands for the Fisher information of a random variable. The above inequalities are easily derived from this general inequality
\begin{eqnarray}
\label{new1}    N(w)J(w) & \geq & 1,
\end{eqnarray}
where the equality holds if and only if $w$ is Gaussian. The inequality (\ref{new1}) is referred to as the isoperimetric inequality for entropies (see \cite{EPI_BrunnInequality:Costa}, \cite{InfInequality:Dembo}, and the references cited therein).

When the variance of $w$ is equal to the variance of $w_{\scriptscriptstyle G}$, the inequality $J(w) \geq J(w_{\scriptscriptstyle G})$ can be derived from $N(w) \leq N(w_{\scriptscriptstyle G})$ using the isoperimetric inequality for entropies. However, we cannot derive the inequality $N(w)\leq N(w_{\scriptscriptstyle G})$ from $J(w) \geq J(w_{\scriptscriptstyle G})$ using the isoperimetric inequality. Instead,  the worst additive noise lemma (see e.g., \cite{EPI:Rioul}, \cite{WAN:Diggavi}, \cite{WAN:Ihara} and the references cited therein) can be derived from the inequality $J(w) \geq J(w_{\scriptscriptstyle G})$ when the variances of $w$ and $w_{\scriptscriptstyle G}$ are identical. All the relationships mentioned above are also valid for random vectors if we substitute either $|\mathbf{J}(\cdot)|^{\frac{1}{n}}$ or $\mathbf{Tr}\{\mathbf{J}(\cdot)\}$ for $J(\cdot)$. The trace and the determinant of a matrix are represented by the notations  $\mathbf{Tr}\{\cdot\}$ and $|\cdot|$, respectively. Since the vector generalization is quite direct, these results  are not mentioned here except the following lemma.

\begin{lemma}[Worst Additive Noise Lemma  \cite{WAN:Diggavi}, \cite{WAN:Ihara}]\label{lem2}
For a random vector $\mathbf{w}$ and a Gaussian random vector $\mathbf{w}_{\scriptscriptstyle G}$ whose covariance matrices are identical to each other,
\begin{eqnarray}
\label{sec2_eq1_1}  I(\mathbf{w}+\mathbf{z}_{\scriptscriptstyle G};\mathbf{z}_{\scriptscriptstyle G}) & \geq & I(\mathbf{w}_{\scriptscriptstyle G}+\mathbf{z}_{\scriptscriptstyle G};\mathbf{z}_{\scriptscriptstyle G}),
\end{eqnarray}
where $I(\cdot;\cdot)$ stands for  mutual information, $\mathbf{z}_{\scriptscriptstyle G}$ is a Gaussian random vector with zero mean and covariance matrix $\boldsymbol{\Sigma}_{\mathbf{z}}$, and all random vectors are independent of one another.
\end{lemma}

Similar to Cram\'{e}r-Rao inequality (see \cite{LargeCRB:Stoica}, \cite{EPI:Rioul}, and the Lemma \ref{lem1}), the worst additive noise lemma shows that the mutual information $I(\mathbf{w}+\mathbf{z}_{\scriptscriptstyle G};\mathbf{z}_{\scriptscriptstyle G})$ is minimized when $\mathbf{w}$ is Gaussian.
Consider that notation $h(\cdot)$ stands for differential entropy, and define the function:
\begin{eqnarray}
\label{sec2_eq2_1}  g(\boldsymbol{\Sigma}_{\mathbf{z}}) = h(\mathbf{w}+\mathbf{z}_{\scriptscriptstyle G})-h(\mathbf{w}_{\scriptscriptstyle G}+\mathbf{z}_{\scriptscriptstyle G})-h(\mathbf{w}) + h(\mathbf{w}_{\scriptscriptstyle G}) .
\end{eqnarray}
The function $g(\cdot)$ is  non-decreasing with respect to the covariance matrix $\boldsymbol{\Sigma}_{\mathbf{z}}$ near the zero matrix $\mathbf{0}$. This is because, due to  Lemma \ref{lem2}, $g(\boldsymbol{\Sigma}_{\mathbf{z}})$ is always non-negative for a covariance matrix $\boldsymbol{\Sigma}_{\mathbf{z}}$ which is arbitrarily close to the zero matrix $\mathbf{0}$. Therefore, near the zero matrix, the first derivative of $g(\boldsymbol{\Sigma}_{\mathbf{z}})$ with respect to $\boldsymbol{\Sigma}_{\mathbf{z}}$ is always positive semi-definite, and using a vector version of de Bruijn's identity \cite{LinVecGaussGradient:Palomar}, the Cram\'{e}r-Rao inequality is derived from the Lemma \ref{lem2} as follows:
\begin{eqnarray}
\label{sec2_eq3_1}  \nabla_{\scriptscriptstyle\boldsymbol{\Sigma}_{\mathbf{z}}} g(\boldsymbol{\Sigma}_{\mathbf{z}}) \Big|_{\boldsymbol{\Sigma}_{\mathbf{z}}=\mathbf{0}} & \succeq & \mathbf{0}\nonumber\\
\Longleftrightarrow \hspace{10mm}\nabla_{\scriptscriptstyle\boldsymbol{\Sigma}_{\mathbf{z}}} I(\mathbf{w}+\mathbf{z}_{\scriptscriptstyle G};\mathbf{z}_{\scriptscriptstyle G})\Big|_{\boldsymbol{\Sigma}_{\mathbf{z}}=\mathbf{0}} - \nabla_{\scriptscriptstyle\boldsymbol{\Sigma}_{\mathbf{z}}}I(\mathbf{w}_{\scriptscriptstyle G}+\mathbf{z}_{\scriptscriptstyle G};\mathbf{z}_{\scriptscriptstyle G})\Big|_{\boldsymbol{\Sigma}_{\mathbf{z}}=\mathbf{0}} & \succeq & \mathbf{0}\nonumber\\
\Longleftrightarrow \hspace{74.5mm}\mathbf{J}(\mathbf{w}) - \mathbf{J}(\mathbf{w}_{\scriptscriptstyle G})& \succeq & \mathbf{0},
\end{eqnarray}
where $\Longleftrightarrow$  stands for  equivalence.

Therefore, in equation (\ref{eq5_3}), the FIM, $\mathbf{J}_{\boldsymbol{\theta}}(\mathbf{y})$, is expressed as
\begin{eqnarray}
\label{sec2_eq4_1}  \mathbf{J}_{\boldsymbol{\theta}}(\mathbf{y}) & = &\mathbb{E}_{\mathbf{x}_{\scriptscriptstyle\boldsymbol{\theta}}} \left[\nabla_{\scriptscriptstyle\boldsymbol{\theta}} \mathbf{x}_{\scriptscriptstyle\boldsymbol{\theta}} \mathbf{J}(\mathbf{w}) \nabla_{\scriptscriptstyle\boldsymbol{\theta}} \mathbf{x}_{\scriptscriptstyle\boldsymbol{\theta}}^T \right]\nonumber\\
& = &2 \mathbb{E}_{\mathbf{x}_{\scriptscriptstyle\boldsymbol{\theta}}} \left[\nabla_{\scriptscriptstyle\boldsymbol{\theta}} \mathbf{x}_{\scriptscriptstyle\boldsymbol{\theta}}   \left(\nabla_{\scriptscriptstyle\boldsymbol{\Sigma}_{\mathbf{z}}} I(\mathbf{w}+\mathbf{z}_{\scriptscriptstyle G};\mathbf{z}_{\scriptscriptstyle G})\Big|_{\boldsymbol{\Sigma}_{\mathbf{z}}=\mathbf{0}}\right) \nabla_{\scriptscriptstyle\boldsymbol{\theta}} \mathbf{x}_{\scriptscriptstyle\boldsymbol{\theta}}^T \right],
\end{eqnarray}
the smallest FIM, $\mathbf{J}_{\boldsymbol{\theta}}(\bar{\mathbf{y}})$, in (\ref{eq12_1}) is expressed as
\begin{eqnarray}
\label{sec2_eq5_1}  \mathbf{J}_{\boldsymbol{\theta}}(\bar{\mathbf{y}})& = & 2 \mathbb{E}_{\mathbf{x}_{\scriptscriptstyle\boldsymbol{\theta}}} \left[\nabla_{\scriptscriptstyle\boldsymbol{\theta}} \mathbf{x}_{\scriptscriptstyle\boldsymbol{\theta}} \left(\nabla_{\scriptscriptstyle\boldsymbol{\Sigma}_{\mathbf{z}}} I(\mathbf{w}_{\scriptscriptstyle G}+\mathbf{z}_{\scriptscriptstyle G};\mathbf{z}_{\scriptscriptstyle G})\Big|_{\boldsymbol{\Sigma}_{\mathbf{z}}=\mathbf{0}}\right)  \nabla_{\scriptscriptstyle\boldsymbol{\theta}} \mathbf{x}_{\scriptscriptstyle\boldsymbol{\theta}}^T \right],
\end{eqnarray}
and
\begin{eqnarray}
\label{sec2_eq6_1}  &&\mathbb{E}_{\mathbf{x}_{\scriptscriptstyle\boldsymbol{\theta}}} \left[\nabla_{\scriptscriptstyle\boldsymbol{\theta}} \mathbf{x}_{\scriptscriptstyle\boldsymbol{\theta}} \left(\nabla_{\scriptscriptstyle\boldsymbol{\Sigma}_{\mathbf{z}}} I(\mathbf{w}+\mathbf{z}_{\scriptscriptstyle G};\mathbf{z}_{\scriptscriptstyle G})\Big|_{\boldsymbol{\Sigma}_{\mathbf{z}}=\mathbf{0}}\right)  \nabla_{\scriptscriptstyle\boldsymbol{\theta}} \mathbf{x}_{\scriptscriptstyle\boldsymbol{\theta}}^T \right]\nonumber\\
& \succeq &
\mathbb{E}_{\mathbf{x}_{\scriptscriptstyle\boldsymbol{\theta}}} \left[\nabla_{\scriptscriptstyle\boldsymbol{\theta}} \mathbf{x}_{\scriptscriptstyle\boldsymbol{\theta}} \left(\nabla_{\scriptscriptstyle\boldsymbol{\Sigma}_{\mathbf{z}}} I(\mathbf{w}_{\scriptscriptstyle G}+\mathbf{z}_{\scriptscriptstyle G};\mathbf{z}_{\scriptscriptstyle G})\Big|_{\boldsymbol{\Sigma}_{\mathbf{z}}=\mathbf{0}}\right) \nabla_{\scriptscriptstyle\boldsymbol{\theta}} \mathbf{x}_{\scriptscriptstyle\boldsymbol{\theta}}^T \right]
\end{eqnarray}

Therefore, one  can do the min-max optimal design based on equations (\ref{sec2_eq4_1}), (\ref{sec2_eq5_1}), and (\ref{sec2_eq6_1}).

\section{PRACTICAL APPLICATIONS}

The min-max approach can be adopted to many applications. One of the typical examples is the  optimal training sequence design for estimating frequency-selective fading channels  \cite{TrainingSeq:Stoica}, \cite{TrainingSeq:Besson}. As a distinctive feature to what was shown in \cite{TrainingSeq:Stoica}, \cite{TrainingSeq:Besson}, the proposed approach does not require neither the assumption of Gaussian noise nor the white noise assumption.

Assume that a linearly modulated signal filtered through a frequency-selective channel is modeled as follows:
\begin{eqnarray}
\label{sec3_eq1_1}  \mathbf{y} & = & \mathbf{X}_{\omega_0} \mathbf{S} \mathbf{h} + \mathbf{w},
\end{eqnarray}
where $\mathbf{y}=[y_0, \cdots, y_{n-1}]^T$, $\mathbf{w}=[w_0, \cdots, w_{n-1}]^T$, $\mathbf{h}=[h_0, \cdots, h_{m-1}]^T$,
\begin{eqnarray}\label{sec3_eq2_1}
\mathbf{X}_{\omega_0} = \left[
\begin{array}{cccc}
    1 & 0 & \cdots & 0 \\
    0 & e^{i \omega_0} & \cdots & 0\\
    \vdots & \cdots & \ddots & \vdots \\
    0 & \cdots & 0 & e^{i (n-1) \omega_0}\\
\end{array}
\right],\quad\quad
\mathbf{S} = \left[
                   \begin{array}{cccc}
                     s_{0} & s_{-1} & \cdots & s_{1-m} \\
                     s_{1} & s_{0} & \cdots & s_{2-m} \\
                     \vdots & \cdots & \ddots & \vdots \\
                     s_{n-1} & s_{n-2} & \cdots & s_{n-m} \\
                   \end{array}
                 \right],
\end{eqnarray}
$\omega_{0}=2\pi f_{0}$ is the frequency offset, $\{s_{1-m}, \ldots, s_{n-1}\}$ stands for the training sequence samples, and $\{h_{0}, \ldots, h_{m-1}\}$ denote  the taps of the channel impulse response, assumed of finite length $m$. The noise $\mathbf{w}$ is an arbitrary random vector with zero mean and noise covariance matrix $\boldsymbol{\Sigma}_{\mathbf{w}}$.

Since we want to find the optimal training sequences to estimate the channel impulse response and the frequency offset, we first define the unknown parameter vector $\boldsymbol{\theta}$ as $[\omega_0, \mathbf{h}_R, \mathbf{h}_I]^T$, where $\mathbf{h}_R$ and $\mathbf{h}_I$ denote the real and the imaginary parts of the channel $\mathbf{h}$.

Based on equation (\ref{eq5_3}),
\begin{eqnarray}
\label{sec3_eq3_1}  \mathbf{J}_{\boldsymbol{\theta}} (\mathbf{y})&= & \mathfrak{Re} \left[\nabla_{\scriptscriptstyle\boldsymbol{\theta}} \boldsymbol{\xi}_{\scriptscriptstyle\boldsymbol{\theta}}  \mathbf{J}(\mathbf{w}) \nabla_{\scriptscriptstyle\boldsymbol{\theta}}\boldsymbol{\xi}_{\scriptscriptstyle\boldsymbol{\theta}}^H \right]\\
\label{sec3_eq3_2}  & \succeq & \mathfrak{Re} \left[\nabla_{\scriptscriptstyle\boldsymbol{\theta}} \boldsymbol{\xi}_{\scriptscriptstyle\boldsymbol{\theta}}   \mathbf{J}(\mathbf{w}_G)  \nabla_{\scriptscriptstyle\boldsymbol{\theta}}\boldsymbol{\xi}_{\scriptscriptstyle\boldsymbol{\theta}}^H \right]\\
\label{sec3_eq3_3}  & \succeq & \mathfrak{Re} \left[\nabla_{\scriptscriptstyle\boldsymbol{\theta}} \boldsymbol{\xi}_{\scriptscriptstyle\boldsymbol{\theta}}   (\lambda_{\min} \mathbf{I})  \nabla_{\scriptscriptstyle\boldsymbol{\theta}}\boldsymbol{\xi}_{\scriptscriptstyle\boldsymbol{\theta}}^H \right]\\
\label{sec3_eq3_4}  & = & \lambda_{\min} \mathfrak{Re} \left[\nabla_{\scriptscriptstyle\boldsymbol{\theta}} \boldsymbol{\xi}_{\scriptscriptstyle\boldsymbol{\theta}} \nabla_{\scriptscriptstyle\boldsymbol{\theta}} \boldsymbol{\xi}_{\scriptscriptstyle\boldsymbol{\theta}}^H \right],
\end{eqnarray}
where $\boldsymbol{\xi}_{\scriptscriptstyle\boldsymbol{\theta}}=\mathbf{X}_{\omega_0} \mathbf{S} \mathbf{h}$, $\lambda_{\min}$ represents the minimum eigenvalue of the FIM, $\mathbf{J}(\mathbf{w}_{\scriptscriptstyle G})$, $\mathfrak{Re}[\cdot]$ denotes the  real part of a vector or  matrix, and superscript $H$ stands for   Hermitian transposition. Since $\boldsymbol{\xi}_{\scriptscriptstyle\boldsymbol{\theta}}$ is a complex-valued function which only depends on the unknown deterministic real parameters, in equation (\ref{sec3_eq3_1}), the equality holds with $\mathfrak{Re}[\cdot]$ and without the expectation. Due to the Lemma \ref{lem1}, equation (\ref{sec3_eq3_2}) is verified, and equation (\ref{sec3_eq3_3}) is satisfied due to  the eigenvalue decomposition.

Equation (\ref{sec3_eq3_4}) reveals  the smallest FIM. It generates the worst CRLB, and it is exactly of the same form as the one shown in \cite{TrainingSeq:Stoica}. Using the same argument as  in \cite{TrainingSeq:Stoica}, the white training sequence is min-max optimal in this case.
This min-max approach heavily depends on how much information we have about the  unknown parameters. If we know the distribution of the noise vector $\mathbf{w}$, then the min-max approach will be adopted based on equation (\ref{sec3_eq3_1}), while equation (\ref{sec3_eq3_2}) will be used when we only know  the covariance matrix of the noise vector $\mathbf{w}$. In both cases, the white training sequences are not optimal since the optimal design is affected by the FIM, $\mathbf{J}(\mathbf{w})$, which is related to the correlation of $\mathbf{w}$. The optimal sequences may depend on either the noise distribution or, at least, the noise covariance matrix. However, without any information about the noise vector $\mathbf{w}$, the white training sequences are  optimal in the sense of minimizing the worst CRLB.

The presented result, i.e.,  for a colored noise $\mathbf{w}$ with given correlation matrix, its FIM $\mathbf{J}_{\boldsymbol{\theta}}(\mathbf{y})$  is minimized when the random vector $\mathbf{w}$ is Gaussian,  can be also interpreted  from a different standpoint as follows. In equation (\ref{eq1_1}), assume $\mathbf{y}$ is passed  through a whitening filter, and a new signal $\tilde{\mathbf{y}}$ is obtained. The noise present  in the new output $\tilde{\mathbf{y}}$ is white since the correlation of the noise is eliminated by the whitening filter. Therefore, we can directly adopt the method proposed in \cite{TrainingSeq:Stoica}. However, the design of the whitening filter requires the covariance matrix of the noise $\mathbf{w}$. If we have information about the covariance matrix of $\mathbf{w}$, we can construct the optimal training sequences; if we do not have information about $\mathbf{w}$, we have to follow the method proposed in equations (\ref{sec3_eq3_3}) and (\ref{sec3_eq3_4}), and use the fact that  the covariance matrix is lower-bounded by the minimum eigenvalue of the covariance matrix multiplied by the identity matrix.

\section{WHAT WE HAVE LEARNED}

The results provided in previous sections show that, given the covariance matrix $\boldsymbol{\Sigma}_{\mathbf{w}}$, the FIM  $\mathbf{J}_{\boldsymbol{\theta}}(\mathbf{y})$, (CRLB) is minimized (respectively maximized) by adopting  the Gaussian assumption. This fact leads  to the min-max optimal approach in the following sense: the FIM  $\mathbf{J_{\boldsymbol{\theta}}(\mathbf{y})}$  (CRLB) depends on the unknown parameters only through the FIM  $\mathbf{J}(\mathbf{w})$. Since the Gaussian noise (not necessarily white) minimizes  the FIM  $\mathbf{J}(\mathbf{w})$, it also minimizes the FIM $\mathbf{J_{\boldsymbol{\theta}}(\mathbf{y})}$ (or equivalently, it maximizes the CRLB). Therefore, the optimal design under the Gaussian assumption yields the best CRLB in the worst case. The CRLB is also expressed using the mutual information. In the information theoretic viewpoint, the fact that a Gaussian random vector minimizes the FIM given the covariance matrix is related to the worst additive noise lemma and the fact that a Gaussian random vector maximizes the differential entropy given the covariance matrix.


%


\section*{ACKNOWLEDGEMENT}

This work was supported by QNRF-NPRP grants 09-391-2-128 and 4-123-2-513.

\section{AUTHORS}
\textit{\textbf{Sangwoo~Park}} (swpark78@neo.tamu.edu) is a Ph.D. student at Texas A\&M University, College Station.

\textbf{\textit{Erchin~Serpedin}} (serpedin@ece.tamu.edu) is a Professor at Texas A\&M University, College Station.

\textit{\textbf{Khalid~Qaraqe}} (khalid.qaraqe@qatar.tamu.edu) is a Professor at Texas A\&M University at Qatar.

\ifCLASSOPTIONcaptionsoff
  \newpage
\fi

\end{document}